\pdfoutput=1
\pdfoutput=1
\documentclass[aps,amsmath,amssymb,preprintnumbers,preprint,nofootinbib,a4paper,11pt]{article}
\usepackage{epsfig}
\usepackage{jheppub,undertilde}
\usepackage{amsthm}
\usepackage{graphicx}
\usepackage{color}
\usepackage{graphicx,textcomp,float,slashed}
\usepackage{arydshln}
\usepackage{empheq}
 
\newcommand{\be}{\begin{equation}}
\newcommand{\ee}{\end{equation}}

\def\bsp#1\esp{\begin{split}#1\end{split}}
\def\bpm{\begin{pmatrix}}
\def\epm{\end{pmatrix}}

\newcommand{\bea}{\begin{eqnarray}}  
\newcommand{\eea}{\end{eqnarray}}  
 \def\bsp#1\esp{\begin{split}#1\end{split}}
\preprint{\begin{flushright} DESY-16-213\\ KCL-PH-TH/2016-63 \end{flushright}}

\title{A Second Higgs Doublet in the Early Universe: \\
Baryogenesis and Gravitational Waves}

\author[1]{G.~C.~Dorsch}
\author[2]{, S.~J.~Huber}
\author[1]{, T.~Konstandin}
\author[2,3]{and J.~M.~No}

\affiliation[1]{DESY, Notkestra\ss e 85, D-22607 Hamburg, Germany}
\affiliation[2]{Department of Physics and Astronomy, University of Sussex, Brighton BN1 9QH, UK}
\affiliation[3]{Department of Physics, King's College London, Strand, WC2R 2LS London, UK}

\abstract{We show that simple Two Higgs Doublet models still provide a 
viable explanation for the matter-antimatter asymmetry of the Universe via electroweak baryogenesis, even after 
taking into account the recent order-of-magnitude improvement on the electron-EDM experimental bound 
by the ACME Collaboration. Moreover we show that, in the region of parameter space where baryogenesis is possible,
the gravitational wave spectrum generated at the end of the electroweak phase transition is within the sensitivity 
reach of the future space-based interferometer LISA.}

\notoc

\begin{document}

\maketitle

\newpage

\section{\Large Introduction}

The origin of the baryon asymmetry of the Universe (BAU) remains one of the most important unsolved puzzles in high 
energy physics and cosmology. Current observations~\cite{WMAP9, Ade:2013zuv} lead to a ratio of the net baryon number per entropy density in the Universe of
\be
	\eta_{\rm obs} \equiv \frac{n_B}{s} \simeq 8.7\times 10^{-11},
\ee
meaning an excess of roughly one baryon for every one billion matter-antimatter annihilation events taking place in the early Universe. 
The three necessary ingredients for generating such an asymmetry dynamically~\cite{Sakharov:1967dj} are 
in principle present within the Standard Model (SM):
{\it (i)} baryon number violation due to the chiral anomaly and non-perturbative sphaleron transitions~\cite{'tHooft:1976up, Jackiw:1976pf, Klinkhamer:1984di}; 
{\it (ii)} violation of charge (C) and charge-parity (CP) symmetries from the electroweak interactions and quark mixings; 
{\it (iii)} displacement from equilibrium coming from the 
Hubble expansion of the Universe and possibly from the electroweak phase transition (EWPT)~\cite{Weinberg:1974hy, Dolan:1973qd, Anderson:1991zb}. A closer 
analysis indicates, however, that the sphalerons and the CP violating diffusion processes are never simultaneously out of equilibrium with respect to the 
Hubble expansion~\cite{Kuzmin:1985mm}, so a BAU can only be generated if the process of electroweak symmetry breaking proceeds via a first order phase transition. Kinetic equilibrium would 
then be broken by the expansion of bubbles of the true vacuum, with a sufficiently large vacuum expectation value (VEV) inside the bubble required in order to avoid washout of the generated asymmetry in the broken phase (see~\cite{Konstandin:2013caa} for a recent review on electroweak baryogenesis). 
As it turns out, this latter condition is \emph{not} satisfied in the SM, since the would-be phase transition is actually a 
smooth crossover~\cite{Kajantie:1996mn, D'Onofrio:2014kta}. Furthermore, a second and unrelated problem is the far too small 
amount of CP violation coming from the CKM matrix, which is suppressed by the Jarlskog invariant~\cite{Jarlskog:1985ht, Agashe:2014kda} 
as well as by the tiny quark Yukawa couplings, hence leading to a prediction for the BAU which is at best ten orders of magnitude below 
the observed value~\cite{Gavela:1993ts,Gavela:1994dt,Huet:1994jb}. 

The BAU is therefore an observable which asks for an extension of the SM with additional sources of CP violation and extra particles coupling 
to the Higgs sector. However, 
the presence of the former has an impact on electric dipole moments (EDMs),
which are tightly constrained experimentally. In particular, there has recently been an update on the electron EDM (eEDM) by the ACME 
collaboration improving the bound by one order of magnitude with respect to the previous experimental limit~\cite{Baron:2013eja}, thus casting doubts 
on whether certain models would still be viable candidates for successful electroweak baryogenesis, and, if so, which regions of their 
parameter space would still be allowed. 
In this work we investigate the current status of baryogenesis in simple 
Two Higgs Doublet Models (2HDMs)\footnote{A recent work has tackled this issue in the context of 2HDM scenarios with an additional 
inert singlet. The presence of an extra singlet tends to strengthen the phase transition and therefore decouples the source of a strong EWPT 
(mainly from the extra singlet) to that of CP violation (which comes from the two doublets), thus alleviating the impact of experimental bounds~\cite{Alanne:2016wtx}.}.
Previous studies on this problem have already established the \emph{a priori} viability of obtaining the BAU 
in this framework~\cite{McLerran:1990zh, Turok:1990zg, Cohen:1991iu, Cline:1996mga, Fromme:2006cm, Cline:2011mm, Shu:2013uua}, but 
only one of them takes the ACME eEDM constraint into account~\cite{Shu:2013uua}, albeit with a parameter set that is now excluded 
by flavor observables. Furthermore, most of these studies assumed a very particular and simplified parameter choice 
for the study of the EWPT (except for~\cite{Cline:2011mm}).
A more recent analysis of the EWPT in 
2HDM scenarios indicates a significantly wider
range of parameters allowing for a strong first order transition~\cite{Dorsch:2013wja,Dorsch:2014qja}, in particular pointing to regions 
of the parameter space with a rather exotic phenomenology so far largely unexplored by collider searches~\cite{Dorsch:2014qja, Dorsch:2016tab}. 
On the other hand, recent analyses on the CP violation front show that the ACME eEDM bound places tight constraints on the CP violating mixing angle among the 
scalars~\cite{Jung:2013hka, Ipek:2013iba, Inoue:2014nva}. Whether the amount of 
allowed CP violation is still sufficient to generate the observed BAU in 2HDMs is a key question we aim to answer in this work.

A first order cosmological phase transition would generate yet another important relic from the early Universe, namely a stochastic gravitational 
wave (GW) background sourced by the dynamics of scalar field bubbles which generates acoustic waves and possibly turbulence in the plasma at 
the very end of the transition. For such a source, active at the electroweak scale, the red-shifted spectrum is expected to peak at 
frequencies $\mathcal{O}(0.1-10$~mHz)~\cite{Kamionkowski:1993fg}, within the range of detectability of 
the near-future space-based GW
interferometer LISA~\cite{AmaroSeoane:2012km}. The importance of observing such a signal cannot be underestimated: it 
would not only provide us with a first image of the early Universe beyond the recombination epoch, but would also constitute an 
alternative, cosmology-based method for probing BSM particle physics which is complementary to collider experiments. The recent 
measurement of GW from binary black hole mergers by LIGO~\cite{Abbott:2016blz, Abbott:2016nmj} has already 
demonstrated our capability to reliably and accurately detect these waves, and has therefore paved the way for using this brand 
new source of information as a probe of physics from cosmological down to microscopic scales.

It is interesting to note that 
the baryon asymmetry and the stochastic GW spectrum resulting from the EWPT behave oppositely as a 
function of the 
expansion velocity of the scalar field bubbles. Baryogenesis is optimal for relatively slow subsonic 
bubble walls, allowing enough time for 
the CP violating diffusion processes to generate an excess of handedness in front of the bubble, later to be converted into a 
BAU by the sphalerons. On the other hand, a detectable stochastic GW spectrum requires a rather strong phase 
transition, releasing a large amount of free-energy which can then be converted into bulk motion of the plasma and kinetic energy 
of the bubbles, thus typically resulting in faster supersonic walls. In particular, when the GW source was modelled as rapidly 
expanding shells of kinetic energy, after the bubble sphericity has been broken by their collision (the so-called ``envelope approximation''), 
then a sizeable spectrum was usually 
predicated on ultra-relativistic walls, in which case electroweak baryogenesis is impossible. However, recent 
developments in the field have significantly improved our understanding of GW generation via acoustic waves~\cite{Hindmarsh:2013xza, Hindmarsh:2015qta}, 
which remain active long after bubble collisions end and are therefore a much more efficient source also in the case of deflagrating 
bubbles\footnote{The impact of turbulence is not yet fully understood, especially the dynamics of its generation from the acoustic waves 
and the efficiency in converting turbulent movement into GWs. Nevertheless, it is known that turbulence can also remain active long after 
the phase transition has completed~\cite{Caprini:2009yp}.}. Moreover, it has been noted that the prospective sensitivity of LISA to power-law 
like spectra can be greatly enhanced by integrating over the frequency of such broadband signals, leading to an improvement of a 
factor $\sim \mathcal{O}(10^3)$ with respect to an estimate based only on the raw sensitivity of the apparatus~\cite{Thrane:2013oya}. 
Using these new developments, we show that the EWPT from 2HDMs could actually lead to \emph{both} an observable BAU and 
detectable GWs by LISA, as a result of yielding rather strong phase transitions with relatively slow moving bubbles.

\section{\Large Two-Higgs-Doublet Models}
\label{section2}
Two Higgs doublet models are among the most minimalistic extensions of the SM, differing from it only by 
the addition of an extra scalar $SU(2)_L$ doublet to its field content. 
In the most general setup the presence of two or more doublets coupling to fermions leads to tree-level flavor changing 
neutral currents, which require some suppression mechanism for agreement with the highly sensitive experimental data. 
We impose here a $\mathbb{Z}_2$ symmetry, forcing each type of fermion to couple to one doublet 
only~\cite{Glashow:1976nt} (see refs.~\cite{Cheng:1987rs,Pich:2009sp,Branco:1996bq, D'Ambrosio:2002ex, Botella:2009pq, Buras:2010mh} for a few alternatives). 
Our focus will be on models of Type II, where leptons and down-type quarks couple to $\Phi_1$ while up-type quarks 
couple to $\Phi_2$~\cite{Hall:1981bc,Branco:2011iw}. If the $\mathbb{Z}_2$-symmetry is exact, however, the scalar 
sector does not break CP, neither explicitly nor spontaneously~\cite{DiazCruz:1992uw}. We therefore allow for soft 
breaking of $\mathbb{Z}_2$, in which case the most general renormalizable and gauge-invariant potential for two doublets can be written as
\be	
   \label{2HDM_potential}
   \begin{split}
	V_{\rm tree}(\Phi_1,\Phi_2)=&
		-\mu^2_1 \Phi_1^{\dagger}\Phi_1-
		\mu^2_2\Phi_2^{\dagger}\Phi_2-
		\frac{1}{2}\left(\mu^2\Phi_1^{\dagger}\Phi_2+{\rm H.c.}\right)+\\
		&+\frac{\lambda_1}{2}\left(\Phi_1^{\dagger}\Phi_1\right)^2+
		\frac{\lambda_2}{2}\left(\Phi_2^{\dagger}\Phi_2\right)^2+
		\lambda_3\left(\Phi_1^{\dagger}\Phi_1\right)\left(\Phi_2^{\dagger}\Phi_2\right)+\\
		 &+\lambda_4\left(\Phi_1^{\dagger}\Phi_2\right)\left(\Phi_2^{\dagger}\Phi_1\right)+
		\frac{1}{2}\left[\lambda_5\left(\Phi_1^{\dagger}\Phi_2\right)^2+{\rm H.c.}\right].
   \end{split}
\ee
Note that $\mu^2$ and $\lambda_5$ can be complex, allowing for explicit CP violation in the scalar sector. 
In this case the VEV of the doublets will also be complex in general, of the form
\be
	\label{vevs}
	\langle\Phi_1\rangle = 
		\frac{1}{\sqrt{2}}
		\begin{pmatrix} 0 \\
				\,v\cos\beta\, 	
		\end{pmatrix}  ,
	\qquad
	\langle\Phi_2\rangle = 
		\frac{1}{\sqrt{2}} 
		\begin{pmatrix} 0 \\
				\,v\sin\beta\,e^{i\theta}\,
		\end{pmatrix},
\ee 
with $v\approx 246.22$~GeV. However, only two of these three complex phases are a priori 
independent~\cite{Davidson:2005cw,Lavoura:1994fv}, since a field redefinition can always be used to 
set one of them to zero. The two field-redefinition-invariant phases can be written as~\cite{Inoue:2014nva}
\be\begin{split}
	\delta_1 &= {\rm Arg}[(\mu^2)^2\lambda_5^*],\\
	\delta_2 &= {\rm Arg}(v_1 v_2^*\, \mu^2 \lambda_5^*).
\end{split}\ee
Moreover, imposing that $V_{\rm tree}$ have a minimum as in eq.~(\ref{vevs}) yields three equations, 
two of which enable us to trade $\mu_1^2$ and $\mu_2^2$ for $v$ and $\tan\beta$, and a third constraining $\delta_1$ and $\delta_2$,
\be
	\label{minimization}
	|\mu^2|\sin(\delta_1 - \delta_2) = v^2\sin\beta \cos\beta\,|\lambda_5|\sin(\delta_1 - 2\delta_2),
\ee
so that there is ultimately only one free CP violating parameter. 
Because the CP violating mixing angle between the three neutral scalars must be small due to EDM constraints, 
it makes sense to speak of two mostly CP-even mass eigenstates, $h^0$ and $H^0$ (with $m_{H^0} \geq m_{h^0}$), and a mostly 
CP-odd state $A^0$ (see {\it e.g.}~\cite{Ipek:2013iba}). A pair of charged scalars $H^\pm$ then completes the scalar spectrum.

We set $m_{h^0}=125$~GeV, identifying the lightest $h^0$ with the Higgs boson observed at 
the LHC~\cite{Aad:2012tfa, Chatrchyan:2012ufa}. A further mixing angle, $\beta-\alpha$, 
regulates how the properties of $h^0$ relate to those of the SM Higgs $h_{\rm SM}$: in the CP 
conserving case, $\beta-\alpha=\pi/2$ corresponds to $h^0=h_{\rm SM}$, the so-called \emph{alignment limit}~\cite{Gunion:2002zf}. 
When CP is violated, this equality can never hold exactly since $h^0$ is not a pure CP-even state. But 
because the allowed CP violating mixings are small, it is still legitimate to speak of alignment, at least to a good approximation.

\subsection{Brief Summary of Experimental Constraints}

Due to the presence of new scalars mediating loop diagrams, oblique corrections to 
electroweak precision observables in 2HDMs~\cite{Grimus:2007if,Grimus:2008nb} (see also~\cite{Gorbahn:2015gxa}) can be quite sizeable, 
particularly affecting the $\rho \equiv m_W^2/m_Z^2\cos^2\theta_W$ parameter. Enforcing $\rho\approx 1$ 
leads to an approximate degeneracy between $H^\pm$ and one of the additional neutral scalars, $H^0$ or $A^0$, 
this being related to the limit when custodial symmetry is approximately recovered~\cite{Haber:2010bw,Funk:2011ad}. 
Moreover, flavor observables whose leading-order contribution in the SM comes from 1-loop diagrams are also highly 
sensitive to the presence of new scalars. In the $\mathbb{Z}_2$-symmetric 2HDM the most important of these 
are $\overline{B}_d-B_d$ mixing and $\overline{B}\to X_s \gamma$ transitions~\cite{Mahmoudi:2009zx,Enomoto:2015wbn}. 
For the latter we use the recent NNLO QCD results from~\cite{Hermann:2012fc,Misiak:2015xwa}. Remarkably, for Type II this 
yields the stringent bound $m_{H^\pm} \geq 480$~GeV at $95$\%~C.L.

CP violating phases are tightly constrained by upper bounds on the neutron and electron EDMs. The relevant effective 
operators are given by~\cite{Engel:2013lsa}
\be\begin{split}
	\mathcal{L} \supset &~
		-\sum_f \frac{d_f}{2}\left( 
			i\bar{f}\sigma_{\mu\nu}\gamma_5 f F^{\mu\nu}
		\right)
		-\sum_f \frac{\widetilde{d}_f}{2}\left(
			i g_s\bar{f}\sigma_{\mu\nu}\gamma_5 f T^aG^{\mu\nu}_a
		\right)\\
		&+ \frac{d_W}{6}f_{abc}\epsilon^{\mu\nu\rho\sigma}G^a_{\mu\lambda}G^{b\ \lambda}_\nu G^c_{\rho\sigma},
\end{split}\ee
with the leading-order contributions to the EDM and chromo-EDM coefficients in 2HDMs, $d_f$ and $\widetilde{d}_f$, coming 
from 2-loop Barr-Zee diagrams, whereas $d_W$ is generated by the 2-loop Weinberg three-gluon operator~\cite{Weinberg:1989dx}. 
Full expressions can be found in refs.~\cite{Inoue:2014nva, Abe:2013qla}. The chromo-EDM and Weinberg operators affect the 
neutron EDM via the running down to the nuclear scale $\Lambda_{\rm QCD}\sim 1$~GeV, which we perform using the 1-loop RGEs for 
the Wilson coefficients~\cite{Dekens:2013zca} and 4-loop QCD running of the strong coupling~\cite{Chetyrkin:1997un}. 

The results are to be compared to the current $90$\% C.L. limits for the electron and neutron EDM. For an illustration of the 
impact of the ACME improved measurement, we also show the constraints from the previous bound coming from experiments done with YbF molecules,
\begin{eqnarray}
|d_e^{\rm ACME}| &< 8.7\times 10^{-29}\ e\cdot{\rm cm},~\text{\cite{Baron:2013eja}}\\
|d_e^{\rm YbF}| &< 1.06\times 10^{-27}\ e\cdot{\rm cm},~\text{\cite{Hudson:2011zz}}\\
|d_n| &< 2.9\times 10^{-26}\ e\cdot {\rm cm}.~\text{\cite{Baker:2006ts}}
\end{eqnarray}
 
 \section{\Large Electroweak Phase Transition and Bubble Wall Velocity}
 \label{sec:PT}

\subsection{Electroweak Phase Transition with a Second Higgs Doublet}

A strong first order EWPT 
(as precisely defined in section~\ref{bubble_Wall_sec}) typically requires large couplings to the scalar particles, which in 
2HDMs translates to sizable splittings among the scalar masses and/or between these masses and the overall (squared) mass 
scale of the second doublet, $M^2\equiv {\rm Re(\mu^2)}/s_{2\beta}$. Now, in order to avoid the decoupling limit 
of the second Higgs doublet or 
instabilities in the scalar potential it is required that\footnote{If $M\gg v$ and the scalar masses are light, some quartic 
couplings will be large (in absolute value) and negative, causing the scalar potential to be unbounded from 
below~\cite{Klimenko:1984qx, Maniatis:2006fs}.} $M\sim v$. On the other hand, if $\langle H^0 \rangle \neq 0$ then $m_{H^0}$ is 
also required to be light in order to avoid a heavy particle getting a VEV and driving the transition, which would tend to reduce 
its strength, as occurs in the SM. 
A relatively heavy $H^0$ remains possible in the 2HDM alignment limit, where
the phase transition is solely driven by $h^0$. In this context a tuned, degenerate 2HDM spectrum 
$m_{H^0}\simeq m_{A^0}\simeq m_{H^\pm} \gg M \sim v$ can still yield a strong 
EWPT\footnote{For such a spectrum unitarity and perturbativity require $\tan\beta \simeq 1$, and any significant departure from this value closes 
the region of a strong EWPT~\cite{DHMN}.}.
This scenario has been studied in~\cite{Fromme:2006cm}, and we will not pursue it further here. 
Still, this highlights that the alignment limit always favours a strong EWPT within the 2HDM, and we will henceforth concentrate on this case 
for simplicity.

Allowing for sizable splittings among the new scalars
significantly enlarges the 2HDM region of parameter space where a strong EWPT is possible~\cite{Dorsch:2013wja, Dorsch:2014qja}. 
Since electroweak precision observables require $H^\pm$ to pair with one of the neutral scalars, and $H^0$ needs to be light 
if 2HDM alignment is only approximate (for a strong EWPT to be viable), it follows that $A^0$ is the only 
scalar which is free to be heavy and induce the required large splittings.
%
%
Thus, a strong EWPT scenario in 2HDMs generically has a hierarchical spectrum, 
with $m_{A^0}-m_{H^0} \gtrsim v$ and $M\sim m_{H^0} \sim v$~\cite{Dorsch:2013wja, Dorsch:2014qja}.

Since flavour observables constrain $m_{H^\pm}>480$~GeV in Type II 2HDM, we choose the pairing $m_{A^0}=m_{H^\pm}$, thus 
arriving at a benchmark scenario with
\begin{center}
	$M=m_{H^0}=200$~GeV, $m_{A^0}=m_{H^\pm}\simeq 480$~GeV $\quad$ (2HDM Type II in alignment).
\end{center}
We note that for $1\leq \tan\beta \leq 5$ the 
quartic couplings are within the perturbativity bound, with max$(\lambda_i) \approx 2\pi$, and tree-level 
unitarity is also satisfied~\cite{Arhrib:2000is,Ginzburg:2003fe,Ginzburg:2005dt}.
We also stress that 2HDM alignment allows for somewhat larger values of $m_{H^0}$ compatible with a strong EWPT, with a similar hierarchical
2HDM spectrum pattern. This hierarchical pattern 
can in fact be probed at the LHC through $A^0\to ZH^0$ searches~\cite{Dorsch:2014qja}, which 
already constrain our above 2HDM 
benchmark scenario to $\tan\beta\gtrsim 1.8$ at 95 \% C.L. from LHC Run 1 data~\cite{Khachatryan:2016are}.

\subsection{Phase Transition Strength \& Bubble Wall Profile}
\label{bubble_Wall_sec}

Since baryogenesis is driven by diffusion processes in front of the bubble wall, we need to 
compute the temperature $T_n$ at which bubble nucleation actually starts, i.e.~at which the probability of nucleating 
one bubble within the Hubble horizon $H^{-1}$ equals unity~\cite{Anderson:1991zb}. This can be obtained straightforwardly 
from the nucleation rate per unit volume~\cite{Linde:1980tt}
\be
	\Gamma/\mathcal{V} \simeq T^4 e^{-S_3/T},
\ee
with $S_3$ the 3-dimensional action of the associated critical bubble. Once bubbles nucleate, they quickly reach a close 
to planar steady state, and their profile can then be approximated by an hyperbolic tangent,
\be
\label{tan_profile}
	\left(\begin{array}{cc} h_1(z) \\ h_2(z) \end{array} \right) = 
		\frac{v_n}{2}\left(\begin{array}{cc} \cos\beta \\ \sin\beta \end{array} \right)
		\left[1-\tanh\left(\frac{z}{L_w}\right)\right],
\ee
where $L_w$ is the wall width and $v_n = \sqrt{h_1^2(T_n) + h_2^2(T_n)}$ is the VEV at the nucleation temperature. 
The phase transition strength is given by the ratio $v_n/T_n$, and one has a strong first order EWPT when $v_n/T_n \geq 1$.
A similar expression to \eqref{tan_profile} holds for the CP violating angle $\theta(z)$ of eq.~(\ref{vevs}), which varies by $\Delta\theta$ along the bubble wall. 

The bubble profile is a saddle point of the action $S_3$, and is therefore computed by solving the corresponding 
equations of motion (EoM) for $h_1(z)$, $h_2(z)$ and $\theta(z)$. While solving this equation is straightforward in a 
one-dimensional case, for which one can use a simple overshooting-undershooting method, in multi-field cases the problem 
becomes much more subtle, because one does not know a priori the path along which the shooting is to be performed. 
Different numerical solutions to this problem have been proposed in the literature~\cite{Kusenko:1995jv, John:1998ip, Cline:1999wi, Konstandin:2006nd}. 
Here we take a two-stepped approach inspired by~\cite{Konstandin:2006nd}. First, a one-dimensional shooting is performed along the path of 
the valley connecting both minima of the scalar potential. The resulting profile is then used as a first approximation to the full solution, 
allowing us to linearize the right hand side of the EoMs by Taylor expanding $\nabla V$. 
The discretized version of the EoMs then becomes a linear system of equations, which can be solved by 
simple (and computationally cheap) matrix inversion. We have verified that the solution obtained from this 
method does satisfy the EoMs. In fact, the first step alone provides a very good approximation to the profile parameters.

A top quark penetrating the bubble wall from the symmetric phase acquires a mass
\be
	m_t(z) = \frac{y_t h_2(z)}{\sqrt{2}} e^{-i\Theta_t(z)},
\ee
and therefore feels the bubble wall as a potential barrier. 
In the semiclassical approximation, the complex phase $\Theta_t$
leads to different dispersion relations for tops and anti-tops, which ultimately induces a non-zero chemical potential 
for left-handed baryons, $\mu_{B_L}$~\cite{Fromme:2006wx}. 
The complex 
phases $\Theta_t$ and $\theta$ are related by~\cite{Alanne:2016wtx, Cline:2011mm, Huber:1999sa}
\be
	\label{tanb_suppression}
	\partial_\mu \Theta_t = 
		-\frac{h_1^2(z)}{h_1^2(z)+h_2^2(z)}\,
		\partial_\mu \theta\,.
\ee
Note that due to this relation, which 
yields $\Delta\Theta_t = -\Delta\theta/(1+\tan^2\beta)$, there is a suppression for $\tan \beta \gg 1$.

The above discussion highlights that the relevant input for the baryon asymmetry computation is the shape of the 
bubble profile, i.e.~the wall thickness $L_w$, the phase transition strength $v_n/T_n$, and the total change in the 
top-quark's CP violating phase, $\Delta\Theta_t$. In principle there is also a dependence on the wall velocity $v_w$, 
expected to be mild as long as the wall remains subsonic~\cite{Fromme:2006cm}. 

\begin{figure}
	\centering
	\includegraphics[width=.47\textwidth]{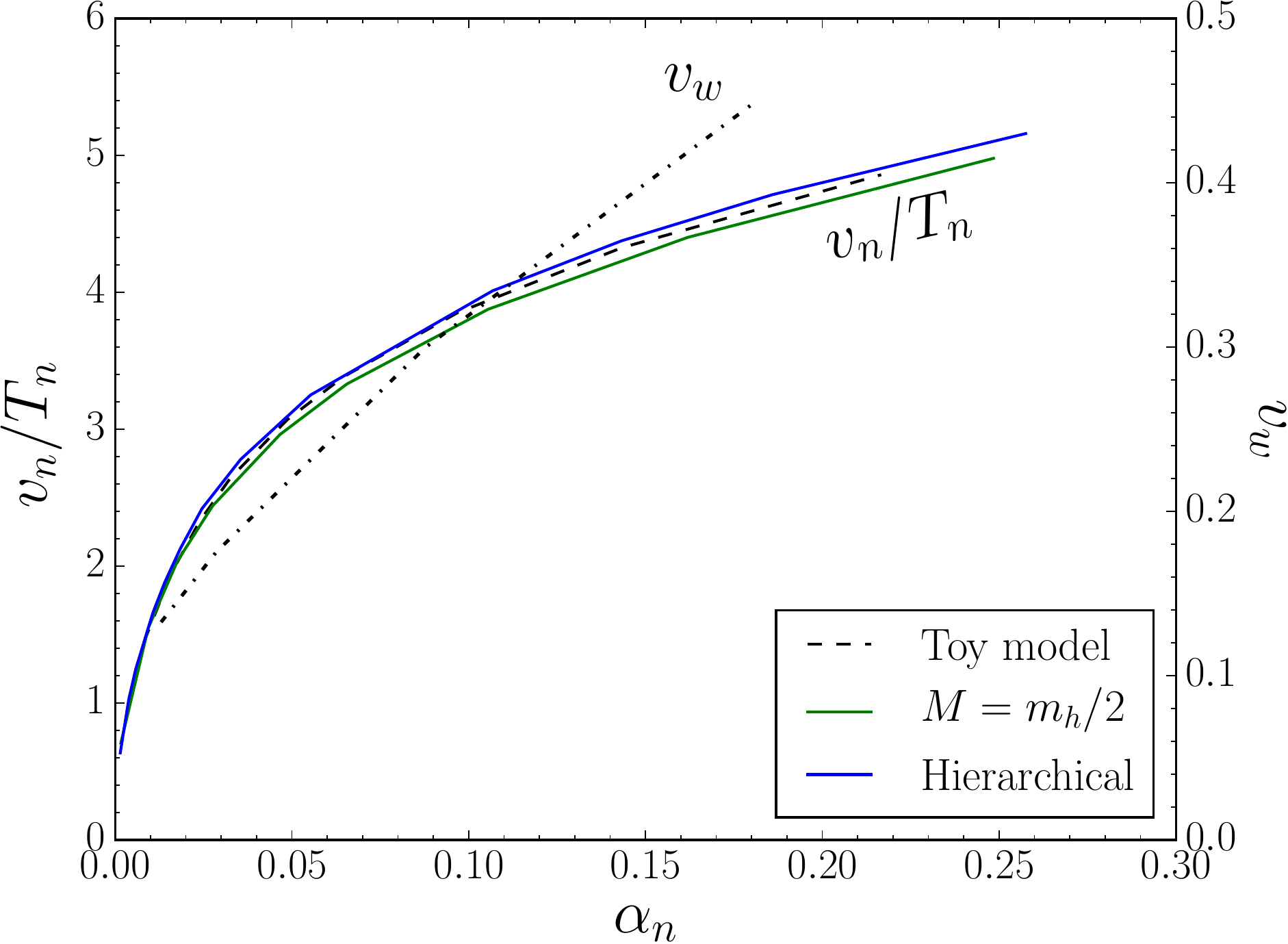}~~~
	\includegraphics[width=.47\textwidth]{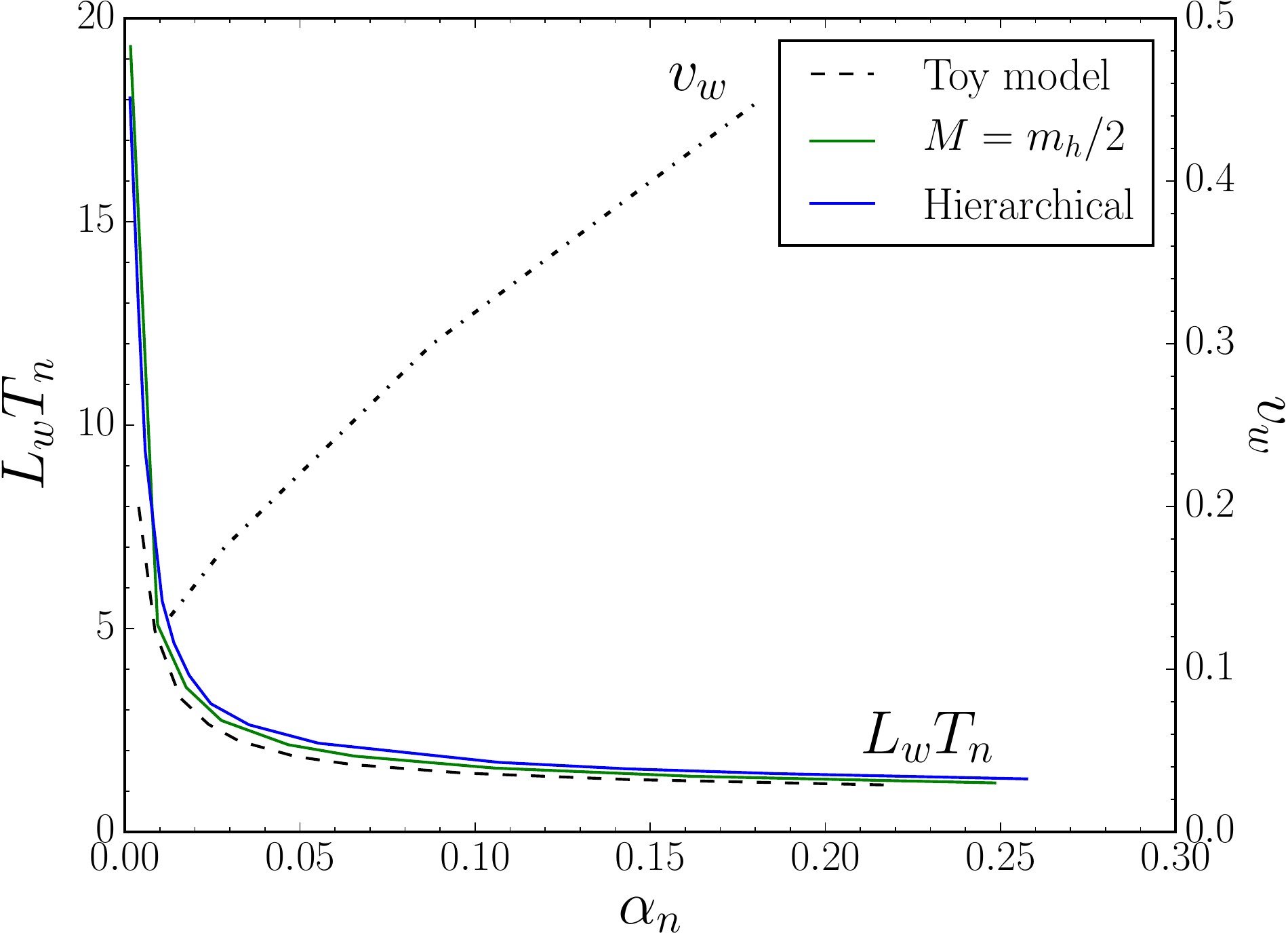}
	\caption{Phase transition strength (left) and wall thickness (right) as a function of fractional vacuum 
	energy released in the plasma at nucleation temperature $T_n$ for the simplified toy model (dashed line), 
	the corresponding 2HDM with $M=m_h/\sqrt{2}$, $\tan\beta=1$ and degenerate masses (green solid line) and the 
	hierarchical case considered throughout this work (blue solid line). Also shown are the wall velocities for the toy model (dot-dashed line).}
	\label{fig:toymodel}
\end{figure}

\subsection{Bubble Wall Velocity}

To estimate $v_w$ we consider a simplified model with four scalars acquiring masses from their 
coupling to a SM-like Higgs according to $m_s = \overline{\lambda} \langle h^0\rangle/\sqrt{2}$. This is equivalent 
to an aligned 2HDM with $M=m_h/\sqrt{2}$ and $\tan\beta=1$, neglecting the self-interactions of the additional scalars. 
This latter simplification, together with the fact that the phase transition dynamics in this toy model involves only one 
scalar field, allows for a more straightforward solution of the EoMs for the scalar field, as is necessary to 
determine the wall velocity. The friction induced by the fluid is modelled by a single friction parameter, $\eta$, 
following~\cite{Huber:2011aa}. In a first step, this parameter is determined at the runaway point~\cite{Bodeker:2009qy,Espinosa:2010hh}, corresponding 
to $\overline{\lambda}=2.29$. It is then extrapolated to weaker transitions by applying the scaling 
$\eta \sim \exp(-\sqrt{v/T})$ found in~\cite{Huber:2011aa}. By construction, this procedure correctly 
reproduces bubble runaway, and leads to a reliable determination of the deflagration/detonation boundary, 
which is crucial for successful baryogenesis. We show in figure~\ref{fig:toymodel} a comparison of the 
relevant phase transition parameters, namely $v_n/T_n$, $L_wT_n$ and the fraction of vacuum energy density 
released in the phase transition in terms of radiation energy in the plasma~\cite{Espinosa:2010hh}, 
\be
	\alpha_n \equiv \frac{\rho_{\rm vac}}{\rho_{\rm rad}},
	\label{alpha}
\ee
for the toy model, the corresponding 2HDM with $M=m_h/\sqrt{2}$, and the hierarchical case considered in the 
rest of the paper. The parameter $\alpha_n$, which will be key for the computation of the GW spectrum 
from the EWPT, can also be seen as a measure of the phase transition strength: the stronger the transition, 
the more energy is released into the plasma, leading to greater $\alpha_n$. Also shown in figure~\ref{fig:toymodel} 
are the values of the wall velocity for the toy model, which show that bubble walls remain subsonic even for very strong 
transitions, $v_n/T_n \sim 4.0$ and $\alpha_n \sim 0.15$. This is the key feature allowing for simultaneous 
baryogenesis and a detectable stochastic GW signal from the EWPT in the 2HDM. 
The good agreement between the shape of the bubble profile 
in the toy model and in the 2HDM, together with the fact that both have the same number of degrees of freedom in 
the plasma with similar couplings, indicates that these values of $v_w$ can also be trusted as estimates for the 
wall velocity in the hierarchical 2HDM considered here.
 
\vspace{-3mm}

\section{\Large Baryogenesis}
\label{sec_baryogenesis}

To compute the baryon asymmetry we use the fluid approximation for the particle distribution functions,
with the chemical potential and the fluid velocity as free-parameters. The corresponding linearized Boltzmann equations
are then solved for the top, anti-top and bottom quarks, the other particles constituting the background~\cite{Bodeker:2004ws}.
The source of displacement from thermal equilibrium as well as of CP violation is the bubble profile, i.e.~the
parameters $v_n/T_n$, $L_w$ and $\Delta\Theta_t$. As discussed in section~\ref{bubble_Wall_sec}, the asymmetric transport of tops and anti-tops
along the bubble wall leads to an excess of handedness in front of the wall, represented by a non-vanishing chemical
potential, $\mu_{B_L}$, for left-handed baryons, to be converted into a baryon asymmetry by the sphalerons (see~\cite{Espinosa:2011eu} for more details).

\begin{table}[h!]\centering
\begin{tabular}{|c|c|c|c|c|c|c|c|}
        \hline
        $m_{A^0}$~[GeV] & $T_n$ & $v_n/T_n$ & $L_w T_n$ & $\Delta\Theta_t$ & $\alpha_n$ & $\beta/H_*$ & $v_w$ \\\hline
        $450$ & $83.665$ & $2.408$ & $3.169$ & $0.0126$ & $0.024$ & $3273.41$ & $0.15$ \\\hline
        $460$ & $76.510$ & $2.770$ & $2.632$ & $0.0083$ & $0.035$ & $2282.42$ & $0.20$\\\hline
        $480$ & $57.756$ & $3.983$ & $1.714$ & $0.0037$ & $0.104$ & $755.62$ & $0.30$ \\\hline
        $483$ & $53.549$ & $4.349$ & $1.556$ & $0.0031$ & $0.140$ & $557.77$ & $0.35$ \\\hline
        $485$ & $50.297$ & $4.668$ & $1.441$ & --- & $0.179$ & $434.80$ & $0.45$ \\\hline
        $487$ & $46.270$ & $5.120$ & $1.309$ & --- & $0.250$ & $306.31$ & $\approx c_s$ \\\hline
\end{tabular}
        \caption{Phase transition parameters relevant for computing the resulting baryon asymmetry (section~\ref{sec_baryogenesis})
        and the gravitational wave spectrum (section~\ref{sec_GWs}), for various pseudoscalar masses $m_{A^0}$ in the hierarchical
        scenario presented in section~\ref{sec:PT}. The values are given for fixed $\tan\beta=2$, but only $\Delta\Theta_t$ is 
        sensitive to $\tan\beta$ according to~\ref{tanb_suppression}.}
        \label{tab:PTdata}
\end{table}

The values of the relevant phase transition parameters entering the computation of the baryon asymmetry are 
shown in Table~\ref{tab:PTdata}, for varying pseudoscalar masses $m_{A^0}$ within the hierarchical 
benchmark discussed in section~\ref{sec:PT}. Notice that for $m_{A^0}\gtrsim 480$~GeV the phase transition is very strong, 
leading to very thin bubble walls, $L_wT_n \sim 1.5$. This may be problematic for the computation of the baryon asymmetry, 
since the formalism of top transport is based on a gradient expansion of the 
Kadanoff-Baym equations~\cite{Prokopec:2003pj, Konstandin:2014zta} with a semi-classical 
treatment of particles in the plasma, such that their momenta $p\sim T_n \gg 1/L_w$~\cite{Cline:1997vk, Cline:2000nw, Kainulainen:2001cn}. 
To account for possible deviations due to our approaching the extreme bound of validity of these 
approximations\footnote{Note that the relevant velocity for baryogenesis is not really $v_w$, but the relative velocity between the bubble wall and the plasma in the deflagration front. The latter may 
be significantly smaller 
than $v_w$~\cite{No:2011fi} in our scenario, yielding a more robust velocity expansion.}, we 
conservatively assume that the unforeseen effects lead to an overestimate of the BAU by a factor $\sim 2$.

\begin{figure}[t!]
	\centering
	\includegraphics[width=0.70\textwidth]{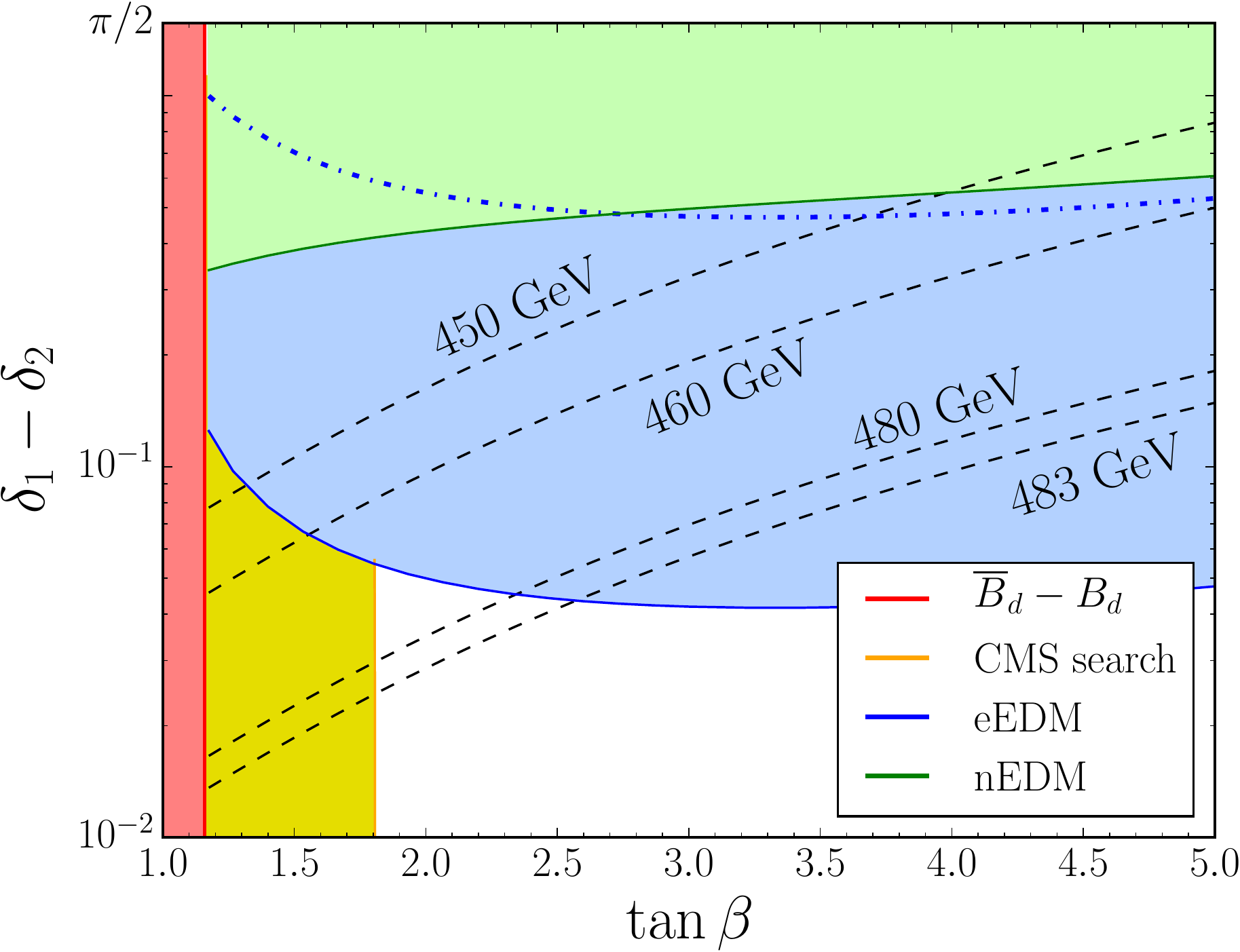}
	\caption{EDM constraints for benchmarks described in text. The dash-dotted line corresponds to the eEDM bound before the ACME experiment. 
	The black dashed lines correspond to the minimum CPV phase necessary for successful baryogenesis for $M=m_{H^0}=200$~GeV and varying $m_{A^0}=m_{H^\pm}$.}
	\label{fig:EDM}
\end{figure}

Figure~\ref{fig:EDM} shows the minimum value of the complex phase $\delta_1-\delta_2$ for which $\eta_B/\eta_{\rm obs}=1$ 
as a function of $\tan\beta$, for $M = m_{H^0} = 200$ GeV and several values of $m_{A^0}=m_{H^\pm}$ 
within the range $[450, 490]$ GeV, 
corresponding to the hierarchical 2HDM benchmark scenario presented in section~\ref{sec:PT}.
As expected, large values of $\tan\beta$ suppress the generation of the BAU due to eq.~(\ref{tanb_suppression}), 
whose effect has to be compensated by a larger value of $\delta_1-\delta_2$ to keep $\eta_B/\eta_{\rm obs}=1$.
The impact of the recent order-of-magnitude improvement on the electron EDM bound from the ACME experiment
is highlighted in figure~\ref{fig:EDM} by showing also the exclusion curve (dotted-dashed blue) from the previous eEDM limit. 
We note that while the neutron-EDM was a competing bound before, the improvement from the ACME experiment now makes the 
eEDM to provide the dominant constraint by far. 
Also shown in figure~\ref{fig:EDM} are the excluded regions from $\overline{B}_d-B_d$ mixing, corresponding to $\tan\beta \lesssim 1.16$, and from 
CMS searches for $A^0\to ZH^0$ with LHC 8 TeV data~\cite{Khachatryan:2016are}, corresponding (for $m_{A^0} = 480$ GeV) to $\tan\beta \lesssim 1.8$.
For $m_{A^0} \approx 480$ GeV there remains then an allowed 
window $1.8\lesssim\tan\beta \lesssim 2.5$ for which the correct BAU could still be obtained in this scenario.
In figure~\ref{fig:EDM} we also present for illustration the results for $m_{A^0}<480$~GeV, potentially excluded by the $\overline{B}\to X_s \gamma$
flavour bound\footnote{This is the case for $m_{A^0}=m_{H^\pm}$. We however note that a small positive mass splitting $m_{H^\pm} - m_{A^0}$ is allowed by 
electroweak precision observables, such as to make the scenario $m_{A^0} \lesssim 480$~GeV potentially compatible with both constraints.}. 
The values of the wall thickness in this case are somewhat larger, $L_wT_n\sim 2-3$, and we can be more confident about the validity of the gradient expansion
(nevertheless the curves shown in figure~\ref{fig:EDM} all take into account the conservative BAU factor $\sim 2$, discussed above, for consistency).
However, for these values of $m_{A^0}$ the bounds from CMS searches are even more stringent, excluding $\tan\beta\lesssim 1.93$, while the eEDM upper bound 
on $\tan\beta$ is also stronger (as a result of a weaker EWPT), altogether closing the baryogenesis window for these masses.

\vspace{1mm}

The discussion above emphasizes that, while baryogenesis is still possible within the 2HDM, quite strong 
phase transitions are required. In fact the usual bound for avoiding sphaleron washout in the broken phase, $v_n/T_n \gtrsim 1.0$, turns out being too mild, 
since EDM constraints alone require significantly stronger transitions if baryogenesis is to be successful.

\vspace{1mm}

Before continuing, let us comment on the fact that similar results could have been obtained for an overall mildly heavier spectrum at the 
cost of tuning. As an example, for $M=m_{H^0}=300$~GeV and $m_{A^0}=m_{H^\pm}\approx 555$~GeV one obtains 
$v_n/T_n = 4.513$, $L_wT_n = 1.625$, $\alpha_n = 0.159$ and $\beta/H_* = 662.85$, values all similar to those of our previously 
considered benchmark with $M=m_{H^0}=200$~GeV and $m_{A^0}=m_{H^\pm}\approx 483$~GeV. 
While the eEDM constraints are hardly affected by this amount of uplifting of the scalar spectrum, 
bounds from CMS $A^0\to ZH^0$ searches get significantly weakened, being currently insensitive to such heavier spectrum. 
Note, however, that an increase in $M=m_{H^0}$ tends to weaken the phase transition, and has to be compensated by 
larger couplings, thus leading to larger values of $\beta/H_*$. As discussed in the next section, even stronger transitions would then be 
required in order to bring this parameter down to the point where the GW spectrum would be observable at LISA, which in turn would lead to 
faster walls, thus harming baryogenesis. 
 
Finally, we stress that in the 2HDM of Type II considered here the Barr-Zee diagrams mediated by top and $W^\pm$ loops interfere 
destructively, with an optimal cancellation for $\tan\beta\sim 1$~\cite{Shu:2013uua, Ipek:2013iba,Inoue:2014nva}, 
leading to milder eEDM constraints in this region as manifestly seen in figure~\ref{fig:EDM}. 
This cancellation does not take place in Type I 2HDM, where the EDM bounds are more severe for low $\tan\beta$, precisely
where baryogenesis is optimal. This shows that accommodating successful baryogenesis in Type I 2HDMs is more challenging.

\section{\Large Gravitational Wave Spectrum}
\label{sec_GWs}
 
While baryogenesis takes place during the period of bubble expansion, gravitational waves start getting sourced at the end of 
the phase transition, when the bubbles collide and overlap. One such source is the uncolliding expanding envelopes of scalar field bubbles, 
since the bubbles' spherical symmetry is broken by their partial overlap. For thermal phase transitions, such as the one we are considering here, 
where the bubble wall has reached a constant velocity long before the bubbles collide, the scalar field contribution is tiny and can be completely 
ignored. Practically the entire energy released by the transition goes into the plasma, as heat and fluid motion. As numerical simulations 
show~\cite{Hindmarsh:2015qta}, the collision of bubbles produces fluid perturbations mostly in the form of sound waves in the plasma, which act as 
a long lived, powerful source of gravitational waves, until they are switched off by the Hubble expansion. For sufficiently strong transitions one 
also expects that the sound waves turn into a stage of turbulence before a Hubble time~\cite{Caprini:2015zlo,Hindmarsh:2016lnk}, with the turbulent 
fluid also acting as a GW source~\cite{Caprini:2009yp}. 

The amplitude of the GW spectrum depends crucially on the amount of energy released in the phase transition and available to be 
converted into GWs, i.e.~the $\alpha$ parameter in eq.~(\ref{alpha}). Another important quantity is the (approximate) inverse duration 
of the phase transition, $\beta$, given in terms of the Hubble rate by
\be
	\frac{\beta}{H_*} = T_*\left.\frac{d(S_3/T)}{dT}\right|_{T_*},
\ee
where $T_*\approx T_n$ is the finalisation temperature at which the phase transition completes\footnote{More precisely, we find that 
typically $T_f \approx 0.96\, T_n$, leading to an approximate $75\%$ difference in $\beta/H$. However, using the finalisation 
temperature actually leads to an overestimate of the average bubble radius during collision and consequently of the GW spectrum. 
We therefore choose to adopt a conservative approach and compute the spectrum at $T_n$.}. Typically, for an electroweak phase 
transition $\beta/H_* \sim \mathcal{O}(100-1000)$. If $\beta$ is large, the bubble nucleation rate increases rapidly with the 
temperature, and the true vacuum then fills the entire space due to bubble nucleation at various different regions. On the other hand, 
a small $\beta$ means that the nucleation rate remains approximately constant for the duration of the phase transition, and space is 
filled by the expansion of the bubbles nucleated at $T_n$. In fact, the bubble radius during collision is $R_* \sim v_w/\beta$, and since 
GWs are sourced by the energy in the moving walls and the accompanying fluid motion, a large signal demands small values of $\beta$.

Because we are focused on deflagrating bubbles, the main source in our case are the sound waves accompanying bubble expansion and 
collision~\cite{Hindmarsh:2013xza, Hindmarsh:2015qta}. This is because the fluid continues to oscillate and source GWs 
even after the transition has completed, leading to an amplitude enhancement by a 
factor $\mathcal{O}\left(\frac{\beta}{H_*}\right)\sim 100-1000$ as compared to the 
spectrum obtained with the envelope approximation. A thorough analytic treatment of this 
case is still lacking (see however~\cite{Hindmarsh:2016lnk}), but numerical simulations indicate 
that the amplitude of the spectrum and its peak frequency can be written as~\cite{Caprini:2015zlo}\footnote{Note that 
shock waves are expected to develop at a time scale $\tau_{\rm sh} \sim \frac{v_w}{\beta}\frac{1}{\sqrt{\kappa_v \alpha}}$~\cite{Caprini:2015zlo,Hindmarsh:2016lnk}, 
which in our case is not necessarily much larger than the lifetime of the acoustic source, $\tau_{\rm sw} \sim H_*^{-1}$~\cite{Hindmarsh:2015qta}, so their 
effects (including some conversion of acoustic energy into vorticity) would have to be taken into account. However, the dynamics of turbulence 
generation from sound waves is still poorly understood, and it is difficult to estimate the impact of this effect on the results presented here. 
We will proceed with the linear sound wave approximation, keeping in mind that more work is needed to fully understand the GW spectrum 
generated from very strong phase transitions such as the ones considered here.} 
\be
	h^2\Omega_{\rm sw} \simeq 2.65\times 10^{-6}\, v_w\,\left(\frac{H_*}{\beta}\right)
						\left(\frac{\kappa_v\alpha}{1+\alpha}\right)^2
						\left(\frac{100}{g_*}\right)^\frac{1}{3},
	\label{OGW}
\ee
\be
	f_{\rm sw} \simeq 1.9\times 10^{-2}\,{\rm mHz}\,\frac{1}{v_w}
					\left(\frac{\beta}{H_*}\right)
					\left(\frac{T}{100\,{\rm GeV}}\right)
					\left(\frac{g_*}{100}\right)^\frac{1}{6}.
	\label{fGW}
\ee
Here $\kappa_v$ is the efficiency in converting the released vacuum energy into bulk motion of the 
fluid, which can be found in ref.~\cite{Espinosa:2010hh} and $g_*\approx 106.75$ 
is the number of relativistic degrees of freedom in the plasma. 

\begin{figure}
	\centering
	\includegraphics[width=.70\textwidth]{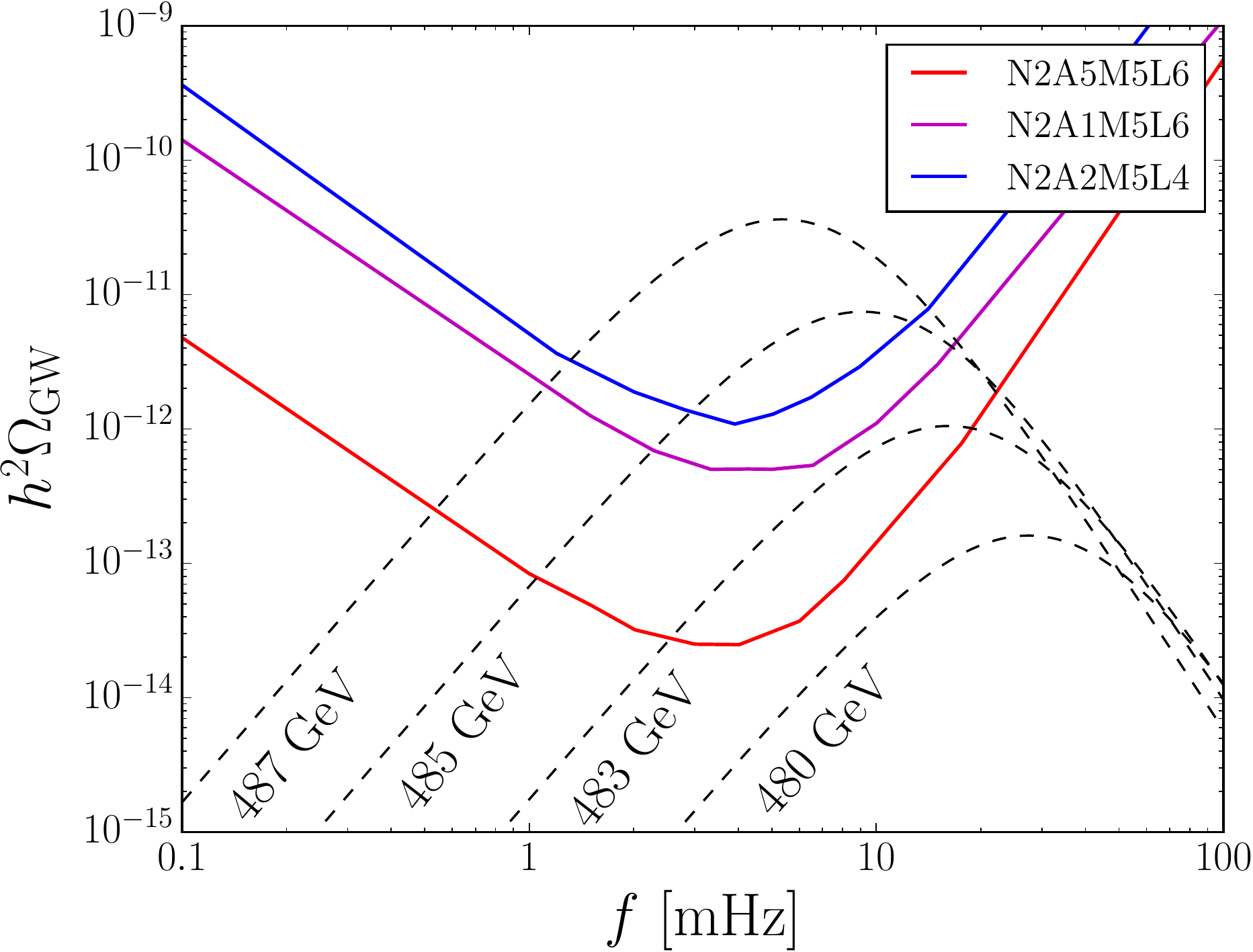}
	\caption{Gravitational wave spectrum for differing values of $m_{A^0}=m_{H^\pm}$. The solid colored lines are the prospective sensitivity for 
	different LISA configurations (see the text and ref.~\cite{Caprini:2015zlo} for more details).} 
	\label{Fig:GW}
\end{figure}

We show in figure~\ref{Fig:GW} the GW spectrum generated by our benchmark scenario with varying values 
for $m_{A^0}=m_{H^\pm}$, with the values of the parameters relevant for obtaining the peak 
amplitude~\eqref{OGW} and frequency~\eqref{fGW} given in Table~\ref{tab:PTdata}. 
Figure~\ref{Fig:GW} also shows the prospective sensitivity for different LISA 
configurations~\cite{Klein:2015hvg, Caprini:2015zlo}. The LISA Pathfinder mission has successfully established the 
noise levels expected for the full experiment (N2), and the configuration with three arms (six links, L6) 
has already been fixed. Thus, the 
remaining free parameters to be determined are the arm lengths (between 1---5~MKm, A1---A5) and the duration of the mission, which we 
set at 5 years (M5). For illustrative purposes we also include the sensitivity curve for two arms (four links, L4) 
with 2~MKm length each (A2). Our results 
are in the same range as those found {\it e.g.} in refs.~\cite{Hashino:2016rvx, Huang:2016cjm} for various other models, provided the phase transition is 
quite strong, as also in our case. 

It is interesting to note that the values for $\beta/H_*$ obtained in the 2HDM are significantly larger 
(for comparable values of $\alpha$) than those usually found in other models 
considered in the GW literature~\cite{Huber:2007vva, Caprini:2015zlo}. This is because $\beta/H_*$ is essentially determined by the temperature 
dependence of the effective potential, which increases with the number of degrees of freedom present in the plasma, as well as with the 
strength of their couplings. Indeed, the hierarchical 2HDM considered here involves relatively strong couplings, 
with the mean field contribution to the thermal potential leading to thermal Higgs masses $m_T^2/T^2 \sim \frac{\lambda_3}{3} \simeq \frac{2\pi}{3}$, larger 
than in weakly coupled scenarios such as supersymmetric extensions.

Finally, we stress that there is some degree of tuning in the results for the GW spectrum, regarding the detectability by LISA. 
For $m_{A^0}=480$~GeV the spectrum is still outside the detectability range of even the most powerful prospective LISA configuration; for $m_{A^0}=487$~GeV 
the walls are already supersonic and no baryogenesis 
would be possible; and for $m_{A^0}\gtrsim 492$~GeV the symmetric vacuum is metastable and no electroweak symmetry breaking takes place.

\section{\Large Conclusions}

We have argued for the possibility that a first order electroweak phase transition could yield the observed baryon asymmetry of the 
Universe and, at the same time, generate a gravitational wave spectrum observable at LISA. 
This may be seen as a ``proof of principle'' for the compatibility of both phenomena, which can coexist 
for rather strong transitions with relatively slow expanding bubbles, as occurs in 2HDM scenarios.
We emphasize that the recent improvements in our understanding of GWs sourced by acoustic 
waves as well as of the prospective LISA sensitivity were vital for the results presented here. In particular, 
although the amplitude enhancement by a factor $\mathcal{O}\left(\beta/{H_*}\right)$ coming from long-lasting sources of GWs has been 
known for a while~\cite{Caprini:2009yp}, a reliable estimate of the dependence of the spectrum with the phase transition parameters and 
the wall velocity, given in~(\ref{OGW}), could only be achieved with very recent data from extensive numerical simulations~\cite{Hindmarsh:2015qta}. 
First steps towards an analytic understanding of the problem have been made in~\cite{Hindmarsh:2016lnk}, but further investigation on the shape of the 
spectrum is granted, especially for very strong phase transitions.

We have also shown that 2HDMs remain viable candidates for explaining the baryon asymmetry of the Universe, even after the recent stringent 
bound on the electron EDM by the ACME collaboration. We note however that the experimental constraints on these scenarios are severe, 
and a significant future increase in the sensitivity of LHC $A^0\to ZH^0$ searches and/or another order-of-magnitude 
improvement of the eEDM bound will rule out the 2HDM in so far as baryogenesis is concerned.

\acknowledgments

We like to thank MIAPP for hospitality and an inspiring atmosphere, where this work started, and we thank 
Germano Nardini and Kimmo Kainulainen for insightful discussions.
G.C.D. and T.K. are supported by the German Science Foundation (DFG) under the Collaborative Research Center (SFB) 
676 Particles, Strings and the Early Universe. The work of S.H is supported by the Science Technology and Facilities 
Council (STFC) under grant number ST/L000504/1.
J.M.N. is partially supported by the People Programme (Marie Curie Actions) of the European Union Seventh Framework Programme (FP7/2007-2013) under REA 
grant agreement PIEF-GA-2013-625809, and by the 
European Research Council under the European Union’s Horizon 2020 program, ERC
Grant Agreement 648680 (DARKHORIZONS)

\vspace{3mm}

\noindent {\bf Note added:} While this paper was being prepared for publication, a similar 
work appeared in the literature claiming the viability of having baryogenesis with a detectable gravitational 
wave spectrum in the context of a singlet extension~\cite{Vaskonen:2016yiu}. We note that the way friction is modelled in that work 
does not seem to lead to a consistent implementation of the runaway phenomenon, thus casting doubt on whether it will lead to a reliable determination of 
the deflagration/detonation boundary, as is crucial for baryogenesis.

\bibliographystyle{JHEP.bst}

\begin{thebibliography}{10}

\bibitem{WMAP9}
{\bf WMAP9} , G.~Hinshaw et~al., ``{\it {Nine-year Wilkinson Microwave
  Anisotropy Probe (WMAP) Observations: Cosmological Parameter Results}}, ''
  {\em The Astrophysical Journal Supplement Series} {\bf 208} (Oct., 2013) 19,
  [\href{http://xxx.lanl.gov/abs/1212.5226}{{\tt arXiv:1212.5226}}].

\bibitem{Ade:2013zuv}
{\bf Planck} , P.~Ade et~al., ``{\it {Planck 2013 results. XVI. Cosmological
  parameters}}, '' {\em Astron.Astrophys.} {\bf 571} (2014) A16,
  [\href{http://xxx.lanl.gov/abs/1303.5076}{{\tt arXiv:1303.5076}}].

\bibitem{Sakharov:1967dj}
A.~Sakharov, ``{\it Violation of {CP} invariance, {C} asymmetry, and baryon
  asymmetry of the {U}niverse}, '' {\em Pisma Zh.Eksp.Teor.Fiz.} {\bf 5} (1967)
  32--35.

\bibitem{'tHooft:1976up}
G.~'t~Hooft, ``{\it {Symmetry breaking through Bell-Jackiw anomalies}}, '' {\em
  Phys.Rev.Lett.} {\bf 37} (1976) 8--11.

\bibitem{Jackiw:1976pf}
R.~Jackiw and C.~Rebbi, ``{\it Vacuum periodicity in a {Y}ang-{M}ills quantum
  theory}, '' {\em Phys.Rev.Lett.} {\bf 37} (1976) 172--175.

\bibitem{Klinkhamer:1984di}
F.~R. Klinkhamer and N.~Manton, ``{\it {A Saddle Point Solution in the
  Weinberg-Salam Theory}}, '' {\em Phys.Rev.} {\bf D30} (1984) 2212.

\bibitem{Weinberg:1974hy}
S.~Weinberg, ``{\it {Gauge and global symmetries at high temperature}}, '' {\em
  Phys.Rev.} {\bf D9} (1974) 3357--3378.

\bibitem{Dolan:1973qd}
L.~Dolan and R.~Jackiw, ``{\it {Symmetry behavior at finite temperature}}, ''
  {\em Phys.Rev.} {\bf D9} (1974) 3320--3341.

\bibitem{Anderson:1991zb}
G.~W. Anderson and L.~J. Hall, ``{\it {The electroweak phase transition and
  baryogenesis}}, '' {\em Phys.Rev.} {\bf D45} (1992) 2685--2698.

\bibitem{Kuzmin:1985mm}
V.~Kuzmin, V.~Rubakov, and M.~Shaposhnikov, ``{\it {On the Anomalous
  Electroweak Baryon Number Nonconservation in the Early Universe}}, '' {\em
  Phys.~Lett.} {\bf B155} (1985) 36.

\bibitem{Konstandin:2013caa}
  T.~Konstandin,
  ``{\it Quantum Transport and Electroweak Baryogenesis},''
  {\em Phys.\ Usp.}\  {\bf 56} (2013) 747
   [{\em Usp.\ Fiz.\ Nauk} {\bf 183} (2013) 785]
  [\href{https://arxiv.org/abs/1302.6713}{\tt arXiv:1302.6713 [hep-ph]}].

\bibitem{Kajantie:1996mn}
K.~Kajantie, M.~Laine, K.~Rummukainen, and M.~E. Shaposhnikov, ``{\it {Is there
  a hot electroweak phase transition at m(H) larger or equal to m(W)?}}, ''
  {\em Phys.Rev.Lett.} {\bf 77} (1996) 2887--2890,
  [\href{http://xxx.lanl.gov/abs/hep-ph/9605288}{{\tt hep-ph/9605288}}].

\bibitem{D'Onofrio:2014kta}
M.~D'Onofrio, K.~Rummukainen, and A.~Tranberg, ``{\it {The Sphaleron Rate in
  the Minimal Standard Model}}, '' {\em Phys.Rev.Lett.} {\bf 113} (2014)
  141602, [\href{http://xxx.lanl.gov/abs/1404.3565}{{\tt arXiv:1404.3565}}].

\bibitem{Jarlskog:1985ht}
C.~Jarlskog, ``{\it {Commutator of the Quark Mass Matrices in the Standard
  Electroweak Model and a Measure of Maximal CP Violation}}, '' {\em
  Phys.Rev.Lett.} {\bf 55} (1985) 1039.

\bibitem{Agashe:2014kda}
{\bf Particle Data Group} , K.~Olive et~al., ``{\it {Review of Particle
  Physics}}, '' {\em Chin.Phys.} {\bf C38} (2014) 090001.

\bibitem{Gavela:1993ts}
M.~Gavela, P.~Hernandez, J.~Orloff, and O.~Pene, ``{\it {Standard model CP
  violation and baryon asymmetry}}, '' {\em Mod.Phys.Lett.} {\bf A9} (1994)
  795--810, [\href{http://xxx.lanl.gov/abs/hep-ph/9312215}{{\tt
  hep-ph/9312215}}].

\bibitem{Gavela:1994dt}
M.~Gavela, P.~Hernandez, J.~Orloff, O.~Pene, and C.~Quimbay, ``{\it {Standard
  model CP violation and baryon asymmetry. Part 2: Finite temperature}}, ''
  {\em Nucl.Phys.} {\bf B430} (1994) 382--426,
  [\href{http://xxx.lanl.gov/abs/hep-ph/9406289}{{\tt hep-ph/9406289}}].

\bibitem{Huet:1994jb}
P.~Huet and E.~Sather, ``{\it {Electroweak baryogenesis and standard model CP
  violation}}, '' {\em Phys.Rev.} {\bf D51} (1995) 379--394,
  [\href{http://xxx.lanl.gov/abs/hep-ph/9404302}{{\tt hep-ph/9404302}}].

\bibitem{Baron:2013eja}
{\bf ACME} , J.~Baron et~al., ``{\it {Order of Magnitude Smaller Limit on the
  Electric Dipole Moment of the Electron}}, '' {\em Science} {\bf 343} (2014)
  269--272, [\href{http://xxx.lanl.gov/abs/1310.7534}{{\tt arXiv:1310.7534}}].

\bibitem{Alanne:2016wtx}
T.~Alanne, K.~Kainulainen, K.~Tuominen, and V.~Vaskonen, ``{\it {Baryogenesis
  in the two doublet and inert singlet extension of the Standard Model}}, ''
  {\em JCAP} {\bf 1608} (2016), no.~08 057,
  [\href{http://xxx.lanl.gov/abs/1607.03303}{{\tt arXiv:1607.03303}}].

\bibitem{McLerran:1990zh}
L.~D. McLerran, M.~E. Shaposhnikov, N.~Turok, and M.~B. Voloshin, ``{\it Why
  the baryon asymmetry of the universe is approximately $10^{-10}$}, '' {\em
  Phys.~Lett.} {\bf B256} (1991) 451--456.

\bibitem{Turok:1990zg}
N.~Turok and J.~Zadrozny, ``{\it {Electroweak baryogenesis in the two doublet
  model}}, '' {\em Nucl.Phys.} {\bf B358} (1991) 471--493.

\bibitem{Cohen:1991iu}
A.~G. Cohen, D.~Kaplan, and A.~Nelson, ``{\it {Spontaneous baryogenesis at the
  weak phase transition}}, '' {\em Phys.~Lett.} {\bf B263} (1991) 86--92.

\bibitem{Cline:1996mga}
J.~M. Cline and P.-A. Lemieux, ``{\it {Electroweak phase transition in two
  Higgs doublet models}}, '' {\em Phys.Rev.} {\bf D55} (1997) 3873--3881,
  [\href{http://xxx.lanl.gov/abs/hep-ph/9609240}{{\tt hep-ph/9609240}}].

\bibitem{Fromme:2006cm}
L.~Fromme, S.~J. Huber, and M.~Seniuch, ``{\it {Baryogenesis in the two-Higgs
  doublet model}}, '' {\em JHEP} {\bf 0611} (2006) 038,
  [\href{http://xxx.lanl.gov/abs/hep-ph/0605242}{{\tt hep-ph/0605242}}].

\bibitem{Cline:2011mm}
J.~M. Cline, K.~Kainulainen, and M.~Trott, ``{\it {Electroweak baryogenesis in
  two Higgs doublet models and B meson anomalies}}, '' {\em JHEP} {\bf 1111}
  (2011) 089, [\href{http://xxx.lanl.gov/abs/1107.3559}{{\tt
  arXiv:1107.3559}}].

\bibitem{Shu:2013uua}
  J.~Shu and Y.~Zhang,
  ``{\it {Impact of a CP Violating Higgs Sector: From LHC to Baryogenesis}}, '' {\em Phys.Rev.Lett.} {\bf 111} (2013), no.~9 091801,
  [\href{http://xxx.lanl.gov/abs/1304.0773}{{\tt arXiv:1304.0773}}].
  
  
  
\bibitem{Dorsch:2013wja}
G.~C. Dorsch, S.~J. Huber, and J.~M. No, ``{\it {A strong electroweak phase
  transition in the 2HDM after LHC8}}, '' {\em JHEP} {\bf 1310} (2013) 029,
  [\href{http://xxx.lanl.gov/abs/1305.6610}{{\tt arXiv:1305.6610}}].

\bibitem{Dorsch:2014qja}
G.~Dorsch, S.~Huber, K.~Mimasu, and J.~No, ``{\it {Echoes of the Electroweak
  Phase Transition: Discovering a second Higgs doublet through $A_0 \rightarrow
  ZH_0$}}, '' {\em Phys.Rev.Lett.} {\bf 113} (2014), no.~21 211802,
  [\href{http://xxx.lanl.gov/abs/1405.5537}{{\tt arXiv:1405.5537}}].

\bibitem{Dorsch:2016tab}
G.~C. Dorsch, S.~J. Huber, K.~Mimasu, and J.~M. No, ``{\it {Hierarchical vs
  Degenerate 2HDM: The LHC Run 1 Legacy at the Onset of Run 2}}, '' {\em Phys.
  Rev.} {\bf D93} (2016), no.~11 115033,
  [\href{http://xxx.lanl.gov/abs/1601.04545}{{\tt arXiv:1601.04545}}].

\bibitem{Jung:2013hka}
M.~Jung and A.~Pich, ``{\it {Electric Dipole Moments in Two-Higgs-Doublet
  Models}}, '' {\em JHEP} {\bf 1404} (2014) 076,
  [\href{http://xxx.lanl.gov/abs/1308.6283}{{\tt arXiv:1308.6283}}].

\bibitem{Ipek:2013iba}
S.~Ipek, ``{\it {Perturbative analysis of the electron electric dipole moment
  and CP violation in two-Higgs-doublet models}}, '' {\em Phys. Rev.} {\bf D89}
  (2014), no.~7 073012, [\href{http://xxx.lanl.gov/abs/1310.6790}{{\tt
  arXiv:1310.6790}}].

\bibitem{Inoue:2014nva}
S.~Inoue, M.~J. Ramsey-Musolf, and Y.~Zhang, ``{\it {CP-violating phenomenology
  of flavor conserving two Higgs doublet models}}, '' {\em Phys. Rev.} {\bf
  D89} (2014), no.~11 115023, [\href{http://xxx.lanl.gov/abs/1403.4257}{{\tt
  arXiv:1403.4257}}].

\bibitem{Kamionkowski:1993fg}
M.~Kamionkowski, A.~Kosowsky, and M.~S. Turner, ``{\it {Gravitational radiation
  from first order phase transitions}}, '' {\em Phys.Rev.} {\bf D49} (1994)
  2837--2851, [\href{http://xxx.lanl.gov/abs/astro-ph/9310044}{{\tt
  astro-ph/9310044}}].

\bibitem{AmaroSeoane:2012km}
P.~Amaro-Seoane et~al., ``{\it {eLISA/NGO: Astrophysics and cosmology in the
  gravitational-wave millihertz regime}}, '' {\em GW Notes} {\bf 6} (2013)
  4--110, [\href{http://xxx.lanl.gov/abs/1201.3621}{{\tt arXiv:1201.3621}}].

\bibitem{Abbott:2016blz}
{\bf Virgo, LIGO Scientific} , B.~P. Abbott et~al., ``{\it {Observation of
  Gravitational Waves from a Binary Black Hole Merger}}, '' {\em Phys. Rev.
  Lett.} {\bf 116} (2016), no.~6 061102,
  [\href{http://xxx.lanl.gov/abs/1602.03837}{{\tt arXiv:1602.03837}}].

\bibitem{Abbott:2016nmj}
{\bf Virgo, LIGO Scientific} , B.~P. Abbott et~al., ``{\it {GW151226:
  Observation of Gravitational Waves from a 22-Solar-Mass Binary Black Hole
  Coalescence}}, '' {\em Phys. Rev. Lett.} {\bf 116} (2016), no.~24 241103,
  [\href{http://xxx.lanl.gov/abs/1606.04855}{{\tt arXiv:1606.04855}}].

\bibitem{Hindmarsh:2013xza}
M.~Hindmarsh, S.~J. Huber, K.~Rummukainen, and D.~J. Weir, ``{\it
  {Gravitational waves from the sound of a first order phase transition}}, ''
  {\em Phys. Rev. Lett.} {\bf 112} (2014) 041301,
  [\href{http://xxx.lanl.gov/abs/1304.2433}{{\tt arXiv:1304.2433}}].

\bibitem{Hindmarsh:2015qta}
M.~Hindmarsh, S.~J. Huber, K.~Rummukainen, and D.~J. Weir, ``{\it {Numerical
  simulations of acoustically generated gravitational waves at a first order
  phase transition}}, '' {\em Phys. Rev.} {\bf D92} (2015), no.~12 123009,
  [\href{http://xxx.lanl.gov/abs/1504.03291}{{\tt arXiv:1504.03291}}].

\bibitem{Caprini:2009yp}
C.~Caprini, R.~Durrer, and G.~Servant, ``{\it {The stochastic gravitational
  wave background from turbulence and magnetic fields generated by a
  first-order phase transition}}, '' {\em JCAP} {\bf 0912} (2009) 024,
  [\href{http://xxx.lanl.gov/abs/0909.0622}{{\tt arXiv:0909.0622}}].

\bibitem{Thrane:2013oya}
E.~Thrane and J.~D. Romano, ``{\it {Sensitivity curves for searches for
  gravitational-wave backgrounds}}, '' {\em Phys. Rev.} {\bf D88} (2013),
  no.~12 124032, [\href{http://xxx.lanl.gov/abs/1310.5300}{{\tt
  arXiv:1310.5300}}].

\bibitem{Glashow:1976nt}
S.~L. Glashow and S.~Weinberg, ``{\it {Natural conservation laws for neutral
  currents}}, '' {\em Phys.Rev.} {\bf D15} (1977) 1958.

\bibitem{Cheng:1987rs}
T.~Cheng and M.~Sher, ``{\it {Mass Matrix Ansatz and Flavor Nonconservation in
  Models with Multiple Higgs Doublets}}, '' {\em Phys.Rev.} {\bf D35} (1987)
  3484.

\bibitem{Pich:2009sp}
A.~Pich and P.~Tuzon, ``{\it {Y}ukawa alignment in the two-{H}iggs-doublet
  model}, '' {\em Phys.Rev.} {\bf D80} (2009) 091702,
  [\href{http://xxx.lanl.gov/abs/0908.1554}{{\tt arXiv:0908.1554}}].

\bibitem{Branco:1996bq}
G.~Branco, W.~Grimus, and L.~Lavoura, ``{\it {Relating the scalar flavor
  changing neutral couplings to the CKM matrix}}, '' {\em Phys.~Lett.} {\bf
  B380} (1996) 119--126, [\href{http://xxx.lanl.gov/abs/hep-ph/9601383}{{\tt
  hep-ph/9601383}}].

\bibitem{D'Ambrosio:2002ex}
G.~D'Ambrosio, G.~Giudice, G.~Isidori, and A.~Strumia, ``{\it {Minimal flavor
  violation: an effective field theory approach}}, '' {\em Nucl.Phys.} {\bf
  B645} (2002) 155--187, [\href{http://xxx.lanl.gov/abs/hep-ph/0207036}{{\tt
  hep-ph/0207036}}].

\bibitem{Botella:2009pq}
F.~Botella, G.~Branco, and M.~Rebelo, ``{\it {Minimal flavour violation and
  multi-Higgs models}}, '' {\em Phys.~Lett.} {\bf B687} (2010) 194--200,
  [\href{http://xxx.lanl.gov/abs/0911.1753}{{\tt arXiv:0911.1753}}].

\bibitem{Buras:2010mh}
A.~J. Buras, M.~V. Carlucci, S.~Gori, and G.~Isidori, ``{\it {Higgs-mediated
  FCNCs: Natural Flavour Conservation vs. Minimal Flavour Violation}}, '' {\em
  JHEP} {\bf 1010} (2010) 009, [\href{http://xxx.lanl.gov/abs/1005.5310}{{\tt
  arXiv:1005.5310}}].

\bibitem{Hall:1981bc}
L.~J. Hall and M.~B. Wise, ``{\it {Flavor changing Higgs-boson couplings}}, ''
  {\em Nucl. Phys.} {\bf B187} (1981) 397--408.

\bibitem{Branco:2011iw}
G.~Branco, P.~Ferreira, L.~Lavoura, M.~Rebelo, M.~Sher, et~al., ``{\it {Theory
  and phenomenology of two-Higgs-doublet models}}, '' {\em Phys.Rept.} {\bf
  516} (Dec., 2012) 1--102, [\href{http://xxx.lanl.gov/abs/1106.0034}{{\tt
  arXiv:1106.0034}}].

\bibitem{DiazCruz:1992uw}
J.~Diaz-Cruz and A.~Mendez, ``{\it Vacuum alignment in multiscalar models}, ''
  {\em Nucl.Phys.} {\bf B380} (1992) 39--50.

\bibitem{Davidson:2005cw}
S.~Davidson and H.~E. Haber, ``{\it {Basis-independent methods for the
  two-Higgs-doublet model}}, '' {\em Phys.Rev.} {\bf D72} (2005) 035004;
  Erratum--ibid. D72 (2005) 099902,
  [\href{http://xxx.lanl.gov/abs/hep-ph/0504050}{{\tt hep-ph/0504050}}].

\bibitem{Lavoura:1994fv}
L.~Lavoura and J.~P. Silva, ``{\it {Fundamental CP violating quantities in a
  SU(2) x U(1) model with many Higgs doublets}}, '' {\em Phys. Rev.} {\bf D50}
  (1994) 4619--4624, [\href{http://xxx.lanl.gov/abs/hep-ph/9404276}{{\tt
  hep-ph/9404276}}].

    
\bibitem{Aad:2012tfa}
{\bf ATLAS Collaboration} , G.~Aad et~al., ``{\it Observation of a new particle
  in the search for the {S}tandard {M}odel {H}iggs boson with the {ATLAS}
  detector at the {LHC}}, '' {\em Phys.~Lett.} {\bf B716} (2012) 1--29,
  [\href{http://xxx.lanl.gov/abs/1207.7214}{{\tt arXiv:1207.7214}}].

\bibitem{Chatrchyan:2012ufa}
{\bf CMS Collaboration} , S.~Chatrchyan et~al., ``{\it {Observation of a new
  boson at a mass of 125 GeV with the CMS experiment at the LHC}}, '' {\em
  Phys.~Lett.} {\bf B716} (2012) 30--61,
  [\href{http://xxx.lanl.gov/abs/1207.7235}{{\tt arXiv:1207.7235}}].

\bibitem{Gunion:2002zf}
  J.~F.~Gunion and H.~E.~Haber, ``{\it {The CP conserving two Higgs doublet model: The Approach to the decoupling limit}}, ''
  {\em Phys.Rev.} {\bf D67} (2003) 075019, [\href{http://xxx.lanl.gov/abs/hep-ph/0207010}{{\tt
  hep-ph/0207010}}].
  
  
  
\bibitem{Grimus:2007if}
W.~Grimus, L.~Lavoura, O.~Ogreid, and P.~Osland, ``{\it {A Precision constraint
  on multi-Higgs-doublet models}}, '' {\em J.Phys.} {\bf G35} (2008) 075001,
  [\href{http://xxx.lanl.gov/abs/0711.4022}{{\tt arXiv:0711.4022}}].

\bibitem{Grimus:2008nb}
W.~Grimus, L.~Lavoura, O.~Ogreid, and P.~Osland, ``{\it {The Oblique parameters
  in multi-Higgs-doublet models}}, '' {\em Nucl.Phys.} {\bf B801} (2008)
  81--96, [\href{http://xxx.lanl.gov/abs/0802.4353}{{\tt arXiv:0802.4353}}].

\bibitem{Gorbahn:2015gxa}
  M.~Gorbahn, J.~M.~No and V.~Sanz,
  ``{\it {Benchmarks for Higgs Effective Theory: Extended Higgs Sectors}}, '' {\em JHEP} {\bf 10} (2015)
  036, [\href{http://xxx.lanl.gov/abs/1502.07352}{{\tt arXiv:1502.07352}}]. 
  
  
\bibitem{Haber:2010bw}
H.~E. Haber and D.~O'Neil, ``{\it {Basis-independent methods for the
  two-Higgs-doublet model III: The CP-conserving limit, custodial symmetry, and
  the oblique parameters S, T, U}}, '' {\em Phys.Rev.} {\bf D83} (2011) 055017,
  [\href{http://xxx.lanl.gov/abs/1011.6188}{{\tt arXiv:1011.6188}}].

\bibitem{Funk:2011ad}
G.~Funk, D.~O'Neil, and R.~M. Winters, ``{\it What the oblique parameters {S},
  {T}, and {U} and their extensions reveal about the 2{HDM}: A numerical
  analysis}, '' {\em Int.J.Mod.Phys.} {\bf A27} (2012) 1250021,
  [\href{http://xxx.lanl.gov/abs/1110.3812}{{\tt arXiv:1110.3812}}].

\bibitem{Mahmoudi:2009zx}
F.~Mahmoudi and O.~Stal, ``{\it {Flavor constraints on the two-Higgs-doublet
  model with general Yukawa couplings}}, '' {\em Phys.Rev.} {\bf D81} (2010)
  035016, [\href{http://xxx.lanl.gov/abs/0907.1791}{{\tt arXiv:0907.1791}}].

\bibitem{Enomoto:2015wbn}
T.~Enomoto and R.~Watanabe, ``{\it {Flavor constraints on the Two Higgs Doublet
  Models of $Z_2$ symmetric and aligned types}}, '' {\em JHEP} {\bf 05} (2016)
  002, [\href{http://xxx.lanl.gov/abs/1511.05066}{{\tt arXiv:1511.05066}}].

\bibitem{Hermann:2012fc}
T.~Hermann, M.~Misiak, and M.~Steinhauser, ``{\it {$\bar{B}\to X_s \gamma$ in
  the Two Higgs Doublet Model up to Next-to-Next-to-Leading Order in QCD}}, ''
  {\em JHEP} {\bf 1211} (2012) 036,
  [\href{http://xxx.lanl.gov/abs/1208.2788}{{\tt arXiv:1208.2788}}].

\bibitem{Misiak:2015xwa}
M.~Misiak et~al., ``{\it {Updated NNLO QCD predictions for the weak radiative
  B-meson decays}}, '' {\em Phys. Rev. Lett.} {\bf 114} (2015), no.~22 221801,
  [\href{http://xxx.lanl.gov/abs/1503.01789}{{\tt arXiv:1503.01789}}].

\bibitem{Engel:2013lsa}
J.~Engel, M.~J. Ramsey-Musolf, and U.~van Kolck, ``{\it {Electric Dipole
  Moments of Nucleons, Nuclei, and Atoms: The Standard Model and Beyond}}, ''
  {\em Prog.Part.Nucl.Phys.} {\bf 71} (2013) 21--74,
  [\href{http://xxx.lanl.gov/abs/1303.2371}{{\tt arXiv:1303.2371}}].

\bibitem{Weinberg:1989dx}
S.~Weinberg, ``{\it {Larger Higgs Exchange Terms in the Neutron Electric Dipole
  Moment}}, '' {\em Phys. Rev. Lett.} {\bf 63} (1989) 2333.

\bibitem{Abe:2013qla}
T.~Abe, J.~Hisano, T.~Kitahara, and K.~Tobioka, ``{\it {Gauge invariant
  Barr-Zee type contributions to fermionic EDMs in the two-Higgs doublet
  models}}, '' \href{http://xxx.lanl.gov/abs/1311.4704}{{\tt arXiv:1311.4704}}.

\bibitem{Dekens:2013zca}
W.~Dekens and J.~de~Vries, ``{\it {Renormalization Group Running of
  Dimension-Six Sources of Parity and Time-Reversal Violation}}, '' {\em JHEP}
  {\bf 1305} (2013) 149, [\href{http://xxx.lanl.gov/abs/1303.3156}{{\tt
  arXiv:1303.3156}}].

\bibitem{Chetyrkin:1997un}
K.~Chetyrkin, B.~A. Kniehl, and M.~Steinhauser, ``{\it {Decoupling relations to
  $\mathcal{O}(\alpha_s^3)$ and their connection to low-energy theorems}}, ''
  {\em Nucl.Phys.} {\bf B510} (1998) 61--87,
  [\href{http://xxx.lanl.gov/abs/hep-ph/9708255}{{\tt hep-ph/9708255}}].

\bibitem{Hudson:2011zz}
J.~J. Hudson, D.~M. Kara, I.~J. Smallman, B.~E. Sauer, M.~R. Tarbutt, and E.~A.
  Hinds, ``{\it {Improved measurement of the shape of the electron}}, '' {\em
  Nature} {\bf 473} (2011) 493--496.

\bibitem{Baker:2006ts}
C.~Baker, D.~Doyle, P.~Geltenbort, K.~Green, M.~van~der Grinten, et~al., ``{\it
  {An Improved experimental limit on the electric dipole moment of the
  neutron}}, '' {\em Phys.Rev.Lett.} {\bf 97} (2006) 131801,
  [\href{http://xxx.lanl.gov/abs/hep-ex/0602020}{{\tt hep-ex/0602020}}].

  

  
\bibitem{Klimenko:1984qx}
K.~Klimenko, ``{\it {On Necessary and Sufficient Conditions for Some Higgs
  Potentials to Be Bounded From Below}}, '' {\em Theor.Math.Phys.} {\bf 62}
  (1985) 58--65.

\bibitem{Maniatis:2006fs}
M.~Maniatis, A.~von Manteuffel, O.~Nachtmann, and F.~Nagel, ``{\it {Stability
  and symmetry breaking in the general two-Higgs-doublet model}}, '' {\em
  Eur.Phys.J.} {\bf C48} (2006) 805--823,
  [\href{http://xxx.lanl.gov/abs/hep-ph/0605184}{{\tt hep-ph/0605184}}].

  
\bibitem{DHMN}
G.~C.~Dorsch, S.~J.~Huber, K.~Mimasu, J.~M.~No, {\it {To Appear}}.
  
  
\bibitem{Arhrib:2000is}
A.~Arhrib, ``{\it {Unitarity constraints on scalar parameters of the standard
  and two Higgs doublets model}}, ''
  \href{http://xxx.lanl.gov/abs/hep-ph/0012353}{{\tt hep-ph/0012353}}.

\bibitem{Ginzburg:2003fe}
I.~Ginzburg and I.~Ivanov, ``{\it {Tree level unitarity constraints in the 2HDM
  with CP violation}}, '' \href{http://xxx.lanl.gov/abs/hep-ph/0312374}{{\tt
  hep-ph/0312374}}.

\bibitem{Ginzburg:2005dt}
I.~Ginzburg and I.~Ivanov, ``{\it {Tree-level unitarity constraints in the most
  general 2HDM}}, '' {\em Phys.Rev.} {\bf D72} (2005) 115010,
  [\href{http://xxx.lanl.gov/abs/hep-ph/0508020}{{\tt hep-ph/0508020}}].

  
\bibitem{Khachatryan:2016are}
  V.~Khachatryan {\it et al.} [CMS Collaboration],
  ``{\it {Search for neutral resonances decaying into a Z boson and a pair of b jets or tau leptons}}, '' {\em Phys. Lett.} {\bf B759} (2016) 369,
  [\href{http://xxx.lanl.gov/abs/1603.02991}{{\tt
  arXiv:1603.02991}}]. 
  
  
\bibitem{Linde:1980tt}
A.~D. Linde, ``{\it {Fate of the false vacuum at finite temperature: theory and
  applications}}, '' {\em Phys.~Lett.} {\bf B100} (1981) 37.

\bibitem{Kusenko:1995jv}
A.~Kusenko, ``{\it {Improved action method for analyzing tunneling in quantum
  field theory}}, '' {\em Phys. Lett.} {\bf B358} (1995) 51--55,
  [\href{http://xxx.lanl.gov/abs/hep-ph/9504418}{{\tt hep-ph/9504418}}].

\bibitem{John:1998ip}
P.~John, ``{\it {Bubble wall profiles with more than one scalar field: A
  Numerical approach}}, '' {\em Phys. Lett.} {\bf B452} (1999) 221--226,
  [\href{http://xxx.lanl.gov/abs/hep-ph/9810499}{{\tt hep-ph/9810499}}].

\bibitem{Cline:1999wi}
J.~M. Cline, G.~D. Moore, and G.~Servant, ``{\it {Was the electroweak phase
  transition preceded by a color broken phase?}}, '' {\em Phys. Rev.} {\bf D60}
  (1999) 105035, [\href{http://xxx.lanl.gov/abs/hep-ph/9902220}{{\tt
  hep-ph/9902220}}].

\bibitem{Konstandin:2006nd}
T.~Konstandin and S.~J. Huber, ``{\it {Numerical approach to multi dimensional
  phase transitions}}, '' {\em JCAP} {\bf 0606} (2006) 021,
  [\href{http://xxx.lanl.gov/abs/hep-ph/0603081}{{\tt hep-ph/0603081}}].

\bibitem{Fromme:2006wx}
L.~Fromme and S.~J. Huber, ``{\it {Top transport in electroweak baryogenesis}},
  '' {\em JHEP} {\bf 0703} (2007) 049,
  [\href{http://xxx.lanl.gov/abs/hep-ph/0604159}{{\tt hep-ph/0604159}}].

\bibitem{Huber:1999sa}
S.~J. Huber, P.~John, M.~Laine, and M.~G. Schmidt, ``{\it {CP violating bubble
  wall profiles}}, '' {\em Phys. Lett.} {\bf B475} (2000) 104--110,
  [\href{http://xxx.lanl.gov/abs/hep-ph/9912278}{{\tt hep-ph/9912278}}].

\bibitem{Huber:2011aa}
S.~J. Huber and M.~Sopena, ``{\it {The bubble wall velocity in the minimal
  supersymmetric light stop scenario}}, '' {\em Phys. Rev.} {\bf D85} (2012)
  103507, [\href{http://xxx.lanl.gov/abs/1112.1888}{{\tt arXiv:1112.1888}}].

\bibitem{Bodeker:2009qy}
  D.~Bodeker and G.~D.~Moore,
  ``{\it {Can electroweak bubble walls run away?}}, '' {\em JCAP} {\bf 0905}
  (2009) 009, [\href{http://xxx.lanl.gov/abs/0903.4099}{{\tt
  arXiv:0903.4099}}].
  
  
\bibitem{Espinosa:2010hh}
J.~R. Espinosa, T.~Konstandin, J.~M. No, and G.~Servant, ``{\it {Energy Budget
  of Cosmological First-order Phase Transitions}}, '' {\em JCAP} {\bf 1006}
  (2010) 028, [\href{http://xxx.lanl.gov/abs/1004.4187}{{\tt
  arXiv:1004.4187}}].

\bibitem{Bodeker:2004ws}
D.~Bodeker, L.~Fromme, S.~J. Huber, and M.~Seniuch, ``{\it {The Baryon
  asymmetry in the standard model with a low cut-off}}, '' {\em JHEP} {\bf 02}
  (2005) 026, [\href{http://xxx.lanl.gov/abs/hep-ph/0412366}{{\tt
  hep-ph/0412366}}].

\bibitem{Espinosa:2011eu}
  J.~R.~Espinosa, B.~Gripaios, T.~Konstandin and F.~Riva,
  ``{\it Electroweak Baryogenesis in Non-minimal Composite Higgs Models},''
  {\em JCAP} {\bf 1201} (2012) 012
  [\href{https://arxiv.org/abs/1110.2876}{\tt arXiv:1110.2876 [hep-ph]}].

\bibitem{Prokopec:2003pj}
T.~Prokopec, M.~G. Schmidt, and S.~Weinstock, ``{\it {Transport equations for
  chiral fermions to order h bar and electroweak baryogenesis. Part 1}}, ''
  {\em Annals Phys.} {\bf 314} (2004) 208--265,
  [\href{http://xxx.lanl.gov/abs/hep-ph/0312110}{{\tt hep-ph/0312110}}].

\bibitem{Konstandin:2014zta}
T.~Konstandin, G.~Nardini, and I.~Rues, ``{\it {From Boltzmann equations to
  steady wall velocities}}, '' {\em JCAP} {\bf 1409} (2014), no.~09 028,
  [\href{http://xxx.lanl.gov/abs/1407.3132}{{\tt arXiv:1407.3132}}].

\bibitem{Cline:1997vk}
J.~M. Cline, M.~Joyce, and K.~Kainulainen, ``{\it {Supersymmetric electroweak
  baryogenesis in the WKB approximation}}, '' {\em Phys. Lett.} {\bf B417}
  (1998) 79--86, [\href{http://xxx.lanl.gov/abs/hep-ph/9708393}{{\tt
  hep-ph/9708393}}]. [Erratum: Phys. Lett.B448,321(1999)].

\bibitem{Cline:2000nw}
J.~M. Cline, M.~Joyce, and K.~Kainulainen, ``{\it {Supersymmetric electroweak
  baryogenesis}}, '' {\em JHEP} {\bf 07} (2000) 018,
  [\href{http://xxx.lanl.gov/abs/hep-ph/0006119}{{\tt hep-ph/0006119}}].

\bibitem{Kainulainen:2001cn}
K.~Kainulainen, T.~Prokopec, M.~G. Schmidt, and S.~Weinstock, ``{\it {First
  principle derivation of semiclassical force for electroweak baryogenesis}},
  '' {\em JHEP} {\bf 06} (2001) 031,
  [\href{http://xxx.lanl.gov/abs/hep-ph/0105295}{{\tt hep-ph/0105295}}].


\bibitem{No:2011fi}
  J.~M.~No,
  ``{\it {Large Gravitational Wave Background Signals in Electroweak Baryogenesis Scenarios}}, '' {\em Phys. Rev.} {\bf D84} (2011)
  124025, [\href{http://xxx.lanl.gov/abs/1103.2159}{{\tt
  arXiv:1103.2159}}]. 
  
  
  
\bibitem{Caprini:2015zlo}
C.~Caprini et~al., ``{\it {Science with the space-based interferometer eLISA.
  II: Gravitational waves from cosmological phase transitions}}, '' {\em JCAP}
  {\bf 1604} (2016), no.~04 001, [\href{http://xxx.lanl.gov/abs/1512.06239}{{\tt
  arXiv:1512.06239}}].

\bibitem{Hindmarsh:2016lnk}
M.~Hindmarsh, ``{\it {Sound shell model for acoustic gravitational wave
  production at a first-order phase transition in the early Universe}}, ''
  \href{http://xxx.lanl.gov/abs/1608.04735}{{\tt arXiv:1608.04735}}.

\bibitem{Klein:2015hvg}
A.~Klein et~al., ``{\it {Science with the space-based interferometer eLISA:
  Supermassive black hole binaries}}, '' {\em Phys. Rev.} {\bf D93} (2016),
  no.~2 024003, [\href{http://xxx.lanl.gov/abs/1511.05581}{{\tt
  arXiv:1511.05581}}].

  
\bibitem{Hashino:2016rvx}
  K.~Hashino, M.~Kakizaki, S.~Kanemura and T.~Matsui,
  Phys.\ Rev.\ D {\bf 94} (2016) no.1,  015005
  [arXiv:1604.02069 [hep-ph]].  
  
  
\bibitem{Huang:2016cjm}
P.~Huang, A.~J. Long, and L.-T. Wang, ``{\it {Probing the Electroweak Phase
  Transition with Higgs Factories and Gravitational Waves}}, ''
  \href{http://xxx.lanl.gov/abs/1608.06619}{{\tt arXiv:1608.06619}}.

\bibitem{Huber:2007vva}
S.~J. Huber and T.~Konstandin, ``{\it {Production of gravitational waves in the
  nMSSM}}, '' {\em JCAP} {\bf 0805} (2008) 017,
  [\href{http://xxx.lanl.gov/abs/0709.2091}{{\tt arXiv:0709.2091}}].

\bibitem{Vaskonen:2016yiu}
V.~Vaskonen, ``{\it {Electroweak baryogenesis and gravitational waves from a
  real scalar singlet}}, '' \href{http://xxx.lanl.gov/abs/1611.02073}{{\tt
  arXiv:1611.02073}}.

\end{thebibliography}

\providecommand{\href}[2]{#2}\begingroup\raggedright\endgroup

 \end{document}